\documentclass[a4paper,superscriptaddress,twocolumn,prl]{revtex4-1}
\usepackage[pdftex]{graphicx}
\usepackage[]{natbib}

\usepackage{float}
\usepackage{epstopdf}
\usepackage{color}
\usepackage{amsmath}
\usepackage[normalem]{ulem}

\begin{document}

\title{High Frequency Sound in a Unitary Fermi Gas}

\author{C. C. N. Kuhn}
\author{S. Hoinka}
\author{I. Herrera}
\author{P. Dyke}
\affiliation{ARC Centre of Excellence in Future Low-Energy Electronics Technologies, Centre for Quantum and Optical Sciences, Swinburne University of Technology, Melbourne 3122, Australia}
\author{J. J. Kinnunen}
\affiliation{Department of Applied Physics, Aalto University, FI-00076 Aalto, Finland}
\author{G. M. Bruun}
\affiliation{Institut for Fysik og Astronomi, Aarhus Universitet, 8000 Aarhus C, Denmark}
\affiliation{Shenzhen Institute for Quantum Science and Engineering and Department of Physics, Southern University of Science and Technology, Shenzhen 518055, China}
\author{C. J. Vale}
\email{cvale@swin.edu.au}
\affiliation{ARC Centre of Excellence in Future Low-Energy Electronics Technologies, Centre for Quantum and Optical Sciences, Swinburne University of Technology, Melbourne 3122, Australia}

\date{\today} 

\begin{abstract}
We present an experimental and theoretical study of the phonon mode in a unitary Fermi gas. Using two-photon Bragg spectroscopy, we measure excitation spectra at a momentum of approximately half the Fermi momentum, both above and below the superfluid critical temperature $T_\mathrm{c}$. Below $T_\mathrm{c}$, the dominant excitation is the Bogoliubov-Anderson (BA) phonon mode, driven by gradients in the phase of the superfluid order parameter. The temperature dependence of the BA phonon is consistent with a theoretical model based on the quasiparticle random phase approximation in which the dominant damping mechanism is via collisions with thermally excited quasiparticles. As the temperature is increased above $T_\mathrm{c}$, the phonon evolves into a strongly damped collisional mode, accompanied by an abrupt increase in spectral width. Our study reveals strong similarities between sound propagation in the unitary Fermi gas and liquid helium. 
\end{abstract}

\maketitle
Elementary excitation spectra provide a quantitative picture of the physical properties of matter. In many-body quantum systems, the lowest lying excitations are typically collective modes such as sound waves, while at higher energies, single-particle excitations dominate. For systems in the normal phase, two distinct regimes of sound propagation generally exist~\cite{Pines_book1999}. When the lifetime $\tau$ of the quasi-particles that comprise the excitation is short compared to the sound frequency $\omega$, i.e. $\omega\tau\ll 1$, local thermodynamic equilibrium can be established and hydrodynamic (first) sound, driven by pressure gradients, is supported. The sound speed is set by the thermodynamic equation of state and the damping depends upon the shear viscosity and thermal conductivity. In the opposite limit, $\omega \tau \gg 1$, the system is said to be collisionless, but a stable linearly dispersing collective mode can exist for repulsive interactions due to a mean-field restoring force, known as zero sound \cite{Pines_book1999}. 
In a superfluid, sound waves are driven by gradients in the phase of the order parameter, as shown by Anderson~\cite{Anderson1958a,Anderson1958b} and Bogoliubov~\cite{Bogoljubov1958}, and later formalised by Goldstone~\cite{Goldstone1961}. At long wavelengths, both the frequency and damping of the Bogoliubov-Anderson (BA) phonon coincide with the hydrodynamic (first) sound mode described by Landau's two-fluid theory \cite{Landau_PR1941,Hohenberg_AP2000,landau2013fluid}. But the BA mode can persist to higher frequencies into the collisionless regime.

Sound modes in strongly interacting gases of ultracold fermions have been investigated both theoretically~\cite{Ohashi_PRA2003, Pieri_PRB2004, Manini_PRA2005, Combescot_PRA2006, Haussmann_PRA2007, Taylor_PRA2009, Braby_PRA2010} and experimentally~\cite{Joseph_PRL2007, Sidorenkov_Nat2013, Tey_PRL2013, Weimer_PRL2015, Hoinka_NatPhys2017, Patel_arXiv2019}. The unitary Fermi gas is of particular interest, since interactions reach the strongest levels allowed by quantum mechanics for a short-range potential, making it an important testing ground for theories of interacting fermions~\cite{BCS-BEC_book2012, Zwierlein_NSchapter2014, Bloch_RMP2008, Ketterle_ebook2007}. Recent quantitative studies of first \cite{Baird_PRL2019, Patel_arXiv2019} and second~\cite{Sidorenkov_Nat2013, Zwierlien_BECtalk} sound propagation have shed light on the transport properties~\cite{Bluhm_PRL2017, Hui_PRA2018} and superfluid fraction of a Fermi gas at unitarity. As the wavevector $k$ is increased beyond the hydrodynamic regime (when $k > 1 / \lambda_{\mathrm{mfp}}$, where $\lambda_{\mathrm{mfp}}$ is the mean free path between collisions), the dispersion of the BA mode remains approximately linear~\cite{Combescot_PRA2006, Kurkjian_PRA2016}, such that the sound speed agrees with the $k \rightarrow 0$ limit. The damping, on the other hand, can show strong departures from hydrodynamic behaviour~\cite{Kurkjian_NJP2017}. Similar features have been seen in neutron scattering studies on liquid helium and these provide a basis for understanding changes in the nature of the elementary excitations across the superfluid transition~\cite{Griffin_book1993, Glyde_RPP2018}.



In this letter, we study sound propagation in a unitary Fermi gas, at a wavevector of approximately half the Fermi wavevector $k_F$, as a function of temperature. Using focused beam Bragg spectroscopy, we map the response of the phonon mode in a region of the cloud with near-homogeneous density~\cite{Hoinka_NatPhys2017}. In the superfluid phase, we observe a clear BA phonon mode in good agreement with a quasiparticle random phase approximation (QRPA) theory. The QRPA theory assumes collisionless dynamics but includes damping via scattering from thermally excited fermionic quasiparticles~\cite{Hoinka_NatPhys2017}. Just above $T_c$, we observe strongly damped collisional sound, which evolves towards single-particle excitations at higher temperatures. Finally, we identify similarities in the temperature dependence of the excitations in the unitary Fermi gas and liquid helium~\cite{Cowley_CJP1971,Glyde_PRB1992,Glyde_RPP2018}


The experimental sequence employed here is similar to that used previously to study the low-temperature excitations in a Fermi gas with tunable interactions \cite{Hoinka_NatPhys2017}. Briefly, we cool a balanced mixture of $^6$Li atoms in the two lowest hyperfine states $|F=1/2,m_\mathrm{F}=\pm1/2\rangle$ and control the temperature of the cloud by varying the endpoint of the evaporation. An external magnetic field is tuned to $832.2$ G, where the $s$-wave scattering length diverges, $|a| \rightarrow \infty$. At this point, elastic collisions are unitarity limited and the thermodynamic properties of the gas become universal functions of the temperature and density \cite{Ho_PRL2004, Nascimbene_Nature2010, landau2013fluid}.
The cloud is held in a harmonic trap with confinement frequencies in the range $\omega_{(x,y,z)}/(2 \pi) = $(115-125, 110-120, 24.5) Hz. A single-frequency 1064 nm laser provides $x$-$y$ confinement, and axial trapping arises from a curvature in the applied magnetic field. The final atom density can be tuned by varying the power in the 1064 nm laser which determines the Fermi energy $E_\mathrm{F} = \hbar^2 k_\mathrm{F}^2 /(2m)= \hbar^2 (3 \pi^2 \bar{n})^{2/3}/2m$, where $\bar{n}$ is the mean density of atoms in the Bragg scattered volume~\cite{Hoinka_NatPhys2017}. 

We use two focused Bragg laser beams with $1/e^2$ radii of 20 $\mathrm{\mu m}$, intersecting at an angle $2\theta=12.9^{\circ}\pm 0.2^{\circ}$, to probe a small volume of the cloud where the density is near-uniform. The Bragg lasers create a periodic perturbation in the centre of the cloud with $k=(4\pi/\lambda)\sin\theta$, where $\lambda = 670$ nm is approximately $1$ THz blue-detuned from the nearest atomic transition. For the atom densities used in subsequent experiments this corresponds to a relative wavevector in the range $0.5 \lesssim k/k_\mathrm{F} \lesssim 0.6$. By scanning the relative frequency of the two Bragg lasers over a range of $\omega/(2 \pi) = 0 \rightarrow \pm 15$ kHz, we map the response of the cloud as a function of the Bragg energy. 
The Bragg lasers are applied with an approximately Gaussian shaped time envelope with a full-width at half-maximum (FWHM) of 600 $\mu$s. Immediately after the Bragg pulse, the optical trap is switched off and the cloud is allowed to expand for 4 ms before taking an absorption image. 
The finite duration and size of the Bragg beams lead to a Fourier-limited spectral resolution of approximately $1.25$ kHz FWHM which is well below the typical Fermi energies, $E_\mathrm{F} \approx 11$ kHz, used in our experiments. 

Within linear response, momentum $P$ is imparted to the cloud at a rate proportional to the imaginary part of the density-density response function $\chi''(k,\omega)$, where $\chi({\bf r-r'},t-t')=-i\vartheta(t-t')\langle[\hat{n}({\bf r},t),\hat{n}({\bf r'},t')]\rangle$~\cite{Brunello_PRA2001}, $\hat{n}({\bf r},t)$ is the density operator at position $\bf{r}$ and time $t$, and $\vartheta(t)$ is the Heaviside function. After a time of flight, the total momentum transferred, $\Delta P$, leads to a displacement of atoms $\Delta x$ from the Bragg scattered volume proportional to $\Delta P$~\cite{Veeravalli_PRL2008}. All measured Bragg spectra are normalized via the $f$-sum rule~\cite{Pines_book1999, KuhnlePRL2010, Hoinka_NatPhys2017},
\begin{align} 
    \frac{\chi''(k,\omega)}{2 \pi n \epsilon_\mathrm{r}} = \frac{\Delta x(k,\omega)}{\int d\omega \, \omega \, \Delta x(k,\omega)},
    \label{chi}
\end{align}
where, $\epsilon_r$ is the atomic recoil energy. 

\begin{figure}[ht]
\centering{}
\includegraphics[width=1\columnwidth]{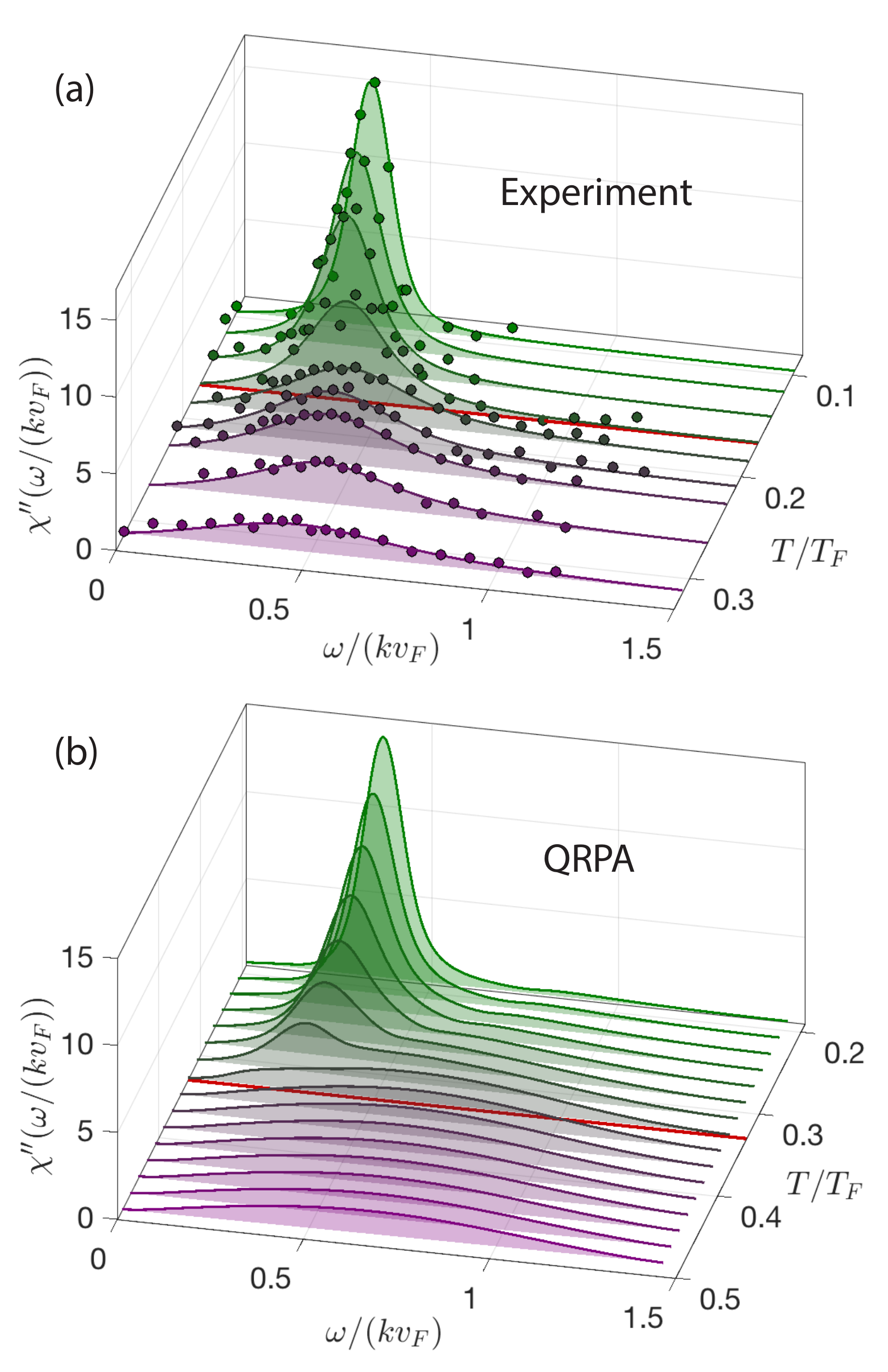}
\caption{(color online) Excitation spectra for a unitary Fermi gas for a range of temperatures $T/T_\mathrm{F}$ at $k/k_\mathrm{F} = 0.55 \pm 0.03$. (a) Filled circles show the experimental data, solid curves and shaded areas are fits to Eq.~\eqref{chi2}, convolved with a Gaussian instrumental broadening function as described in the text. (b) Calculated excitation spectra using the QRPA theory, also convolved with the instrumental broadening function. Red solid lines indicate the superfluid transition temperature $T_c$. All spectra are presented in units of $(4\pi \epsilon_\mathrm{r} E_\mathrm{F})^{-1}$.}
\label{spectra}
\end{figure}

Figure \ref{spectra}(a) shows a selection of measured spectra at unitarity for a range of temperatures below and above $T_\mathrm{c} \simeq 0.17 \; T_\mathrm{F}$. Cloud temperatures were determined by fitting the \emph{in-situ} density profiles to the equation of state \cite{Ku_Science2012}. Below $T_\mathrm{c}$, we observe a well-defined phonon mode with a decreasing amplitude and increasing width as the temperature approaches $T_\mathrm{c}$. Above $T_\mathrm{c}$, the mode is noticeably broader, becoming strongly damped with increasing temperature. 

Figure \ref{spectra}(b) shows calculated spectra using the QRPA theory~\cite{Hoinka_NatPhys2017}. Note that the QRPA theory yields a different $T_\mathrm{c} \sim 0.33\;T_\mathrm{F}$ to the experiment, but the behaviour as a function of $T/T_\mathrm{c}$ can be compared. The evolution of the BA mode in the calculated spectra shows good agreement with the experimental data below $T_\mathrm{c}$. The theory assumes collisionless dynamics and predicts a well-defined BA mode in the superfluid phase, but no collective mode in the normal phase due to the absence of zero sound for attractive interactions \cite{Pines_book1999}. Above $T_\mathrm{c}$, the QRPA spectra consist of just single-particle excitations. 

Inspired by studies of the excitations in liquid helium, we fit our experimental spectra to a damped harmonic oscillator function~\cite{Cowley_CJP1971,Glyde_PRB1992,Glyde_RPP2018}
\begin{align} 
    \chi''(k,\omega) = \frac{  Z \, k \Gamma \omega} {(\omega^2-\omega^2_0)^2 + \omega^2\Gamma^2}, 
    \label{chi2}
\end{align}
where $\Gamma$ is the damping rate, $\omega_0$ is the fundamental frequency of the mode, and $Z$ is a normalisation factor. Equation \eqref{chi2} is further convolved with a Gaussian broadening function with FWHM 1.25 kHz before performing the fit, to account for the finite spectral resolution of our measurements. Solid lines and shaded areas in Fig.~\ref{spectra}(a) show the fitted functions.

\begin{figure}[ht]
\includegraphics[width=1\columnwidth]{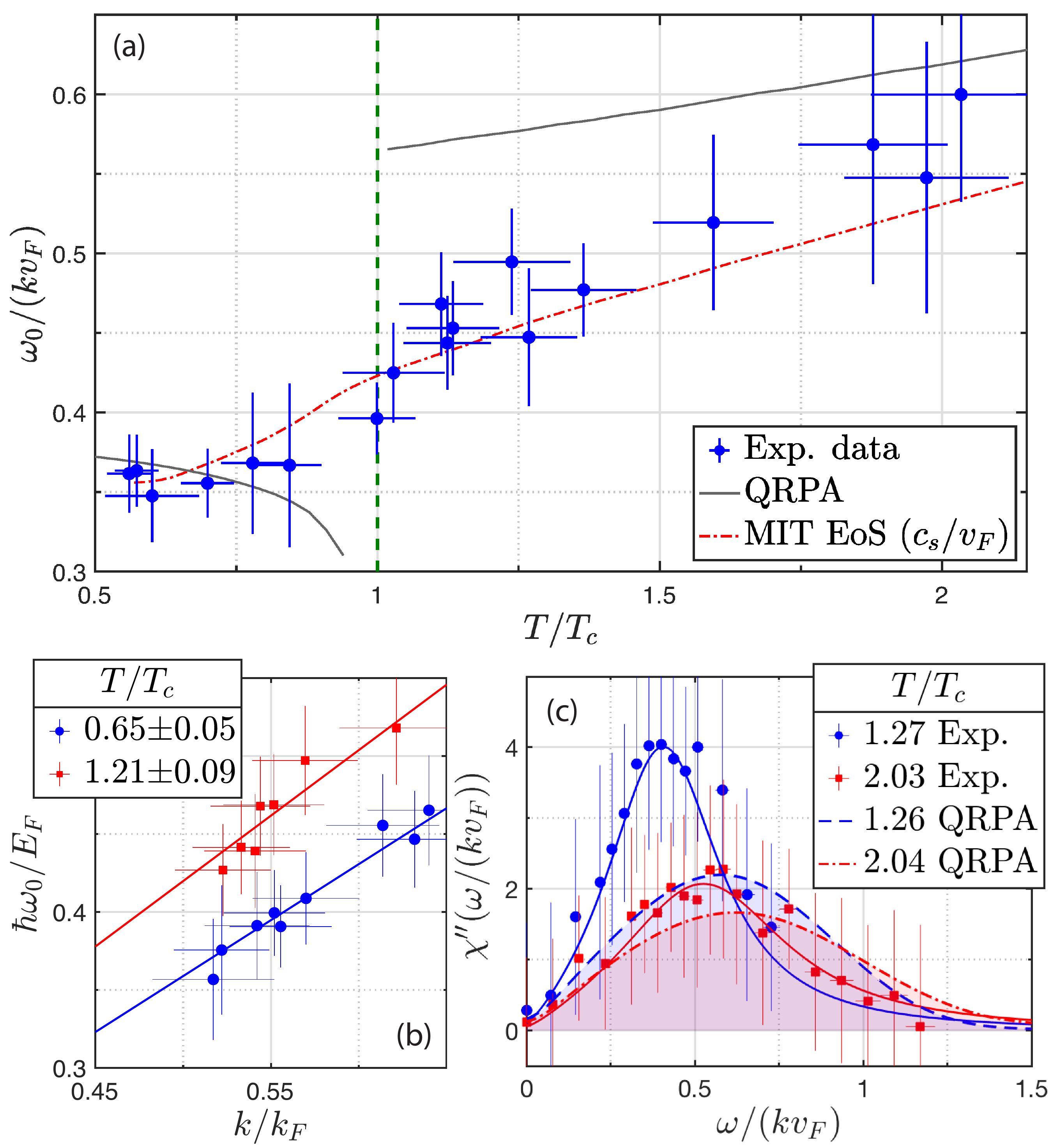}
\caption{(color online) (a) Mode frequency as function of $T/T_\mathrm{c}$. Blue circles are experimental data, the dash-dotted red line is the hydrodynamic sound frequency \cite{Ku_Science2012}, and the grey solid line is the QRPA prediction. $T_\mathrm{c}$ is indicated by the green dashed line. (b) Mode frequency as a function of momentum at two temperatures, below and above $T_c$, respectively. Straight lines are linear fits that cross the origin confirming the dispersion is linear regime within experimental resolution. (c) shows zoomed in plots of two experimental spectra (points and solid lines) in the normal phase at two different temperatures along with the corresponding QRPA spectra (dashed and dash-dotted shaded curves). Error bars include uncertainties in the density measurement used to determine $E_\mathrm{F}$ and statistical (shot-to-shot) fluctuations in the measured Bragg response.}
\label{Speed_sound}
\end{figure}

Blue points in Fig. \ref{Speed_sound}(a) shows the measured centre frequency $\omega_0$ of the spectra which, provided the dispersion is approximately linear, serves as a measure of the sound speed. In Fig. \ref{Speed_sound}(b) we plot $\omega_0$ over a small range of $k/k_\mathrm{F}$ at temperatures below and above $T_\mathrm{c}$, $T/T_\mathrm{F} = 0.11(2)$ (blue circles) and $T/T_\mathrm{F} = 0.20(2)$, (red squares), respectively. Fitting these data to $\hbar \omega_0/E_\mathrm{F} = a (k/k_\mathrm{F})^b$, where $a$ and $b$ are fitting parameters, we find $b=1.1\pm 0.2$ and $1.1 \pm 0.48$, at the low and high temperatures, respectively, indicating that the observed dispersion is linear (sonic) within our experimental resolution \cite{Combescot_PRA2006, Kurkjian_PRA2016, Bighin_PRA2015}.

Also shown in Fig. \ref{Speed_sound}(a) are the mode frequencies from the QRPA calculation (grey solid line), and, the hydrodynamic sound speed obtained from the thermodynamic equation of state \cite{Ku_Science2012} using, $c_\mathrm{s}^2 = (\partial {\cal{P}} / \partial \rho)_s$, where $\cal{P}$ is the pressure, $\rho=n m$ is the mass density and $s$ is the entropy density (red dash-dotted line). The experimental data for the sound speed are consistent with both the hydrodynamic result and the QRPA theory for $T < T_\mathrm{c}$. As $T$ approaches $T_\mathrm{c}$ from below, the QRPA theory predicts a downward shift of the BA mode frequency, which is ``repelled'' from the single-particle continuum. The threshold for single-particle excitations is set by twice the superfluid gap $2\Delta$ \cite{Combescot_PRA2006, Hoinka_NatPhys2017}, which decreases with temperature. However This downshift is not visible in the experimental data. Above $T_\mathrm{c}$, our data agree well with the hydrodynamic sound speed, but lies well below the collisionless single-particle calculation. Fig. \ref{Speed_sound}(c) shows a comparison of two sets of experimental and theoretical spectra, in the normal phase at temperatures close to, and well above, $T_\mathrm{c}$. While the single particle spectrum just above is very wide just above $T_\mathrm{c}$, the experimental data show a much narrower peak, consistent with a damped collisional mode in the normal phase. Further above $T_\mathrm{c}$, the agreement improves, reflecting the evolution towards single particle excitations at higher temperatures. 
A Boltzmann equation calculation of the viscous relaxation time $\tau$ supports this, yielding $0.5 \lesssim \omega_0 \tau \lesssim 1$ for $T\gtrsim T_\mathrm{c}$ \cite{Massignan2005,Bruun2005} with $\tau$ growing monotonically with increasing $T$. Similar values of $\omega_0\tau$ were obtained using a Luttinger-Ward \cite{Enss_AP2011} and Gaussian pair fluctuation theories \cite{Hui_PRA2018}. Our measurements above $T_\mathrm{c}$, thus lie in the crossover between the hydrodynamic and collisionless regimes. 

 \begin{figure*}[ht]
  \includegraphics[width=\textwidth]{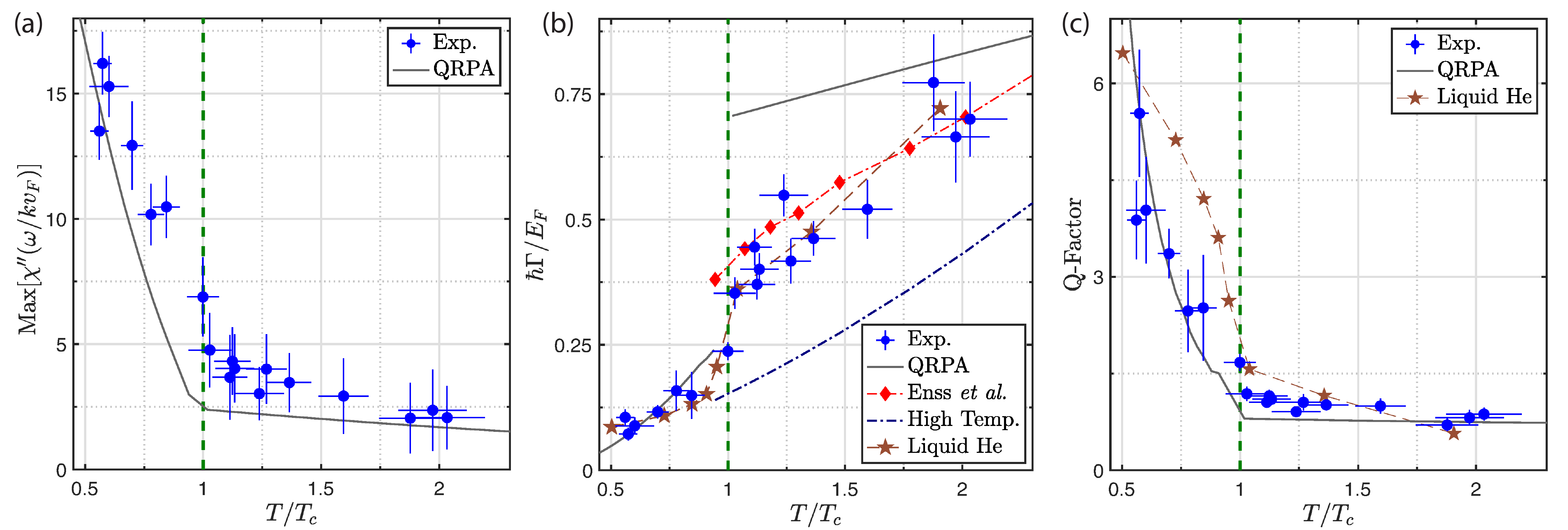}
\caption{(color online) (a) Normalised peak amplitude of the phonon mode as function of temperature. Blue circles are experimental data and the gray solid line is the QRPA calculation. (b) Blue points show the FWHM of the experimental spectra determined by fitting Eq. \eqref{chi2}. Grey solid line is the damping predicted by the QRPA theory, blue  and red dash-dotted lines are hydrodynamic predictions from Eq.~\eqref{eta} using, the classical result for the shear viscosity \cite{Bruun2005} and for $\eta$ obtained from a ${\mathcal T}$-matrix calculation~\cite{Enss_AP2011}. (c) Quality factor $\omega_0/\Gamma$ of the sound mode. Blue circles are the experimental data, grey solid line is the QRPA theory. Brown stars and dashed lines in (b) and (c) show scaled results for liquid helium measured at a wavevector of $0.4 \, \mathring{\rm{A}}^{-1}$~\cite{Cowley_CJP1971}.}
\label{FWHM}
\end{figure*}

In Fig.~\ref{FWHM}, we plot the amplitude $\text{Max}[\chi''(k,\omega)]$, width $\Gamma$, and quality factor $Q = \omega_0/\Gamma$, of the phonon mode as a function of the temperature. The amplitude, \ref{FWHM}(a), decreases monotonically with $T$ with a steep gradient below $T_\mathrm{c}$ and shallower gradient above $T_\mathrm{c}$. The width, Fig.~\ref{FWHM}(b), increases with temperature and exhibits a jump at the superfluid transition reflecting a sudden increase in the damping. These features signify a qualitative change in the sound mode from a BA phonon below $T_\mathrm{c}$ to a strongly damped collisional mode in the normal phase.


The QRPA theory accurately describes both the amplitude and width below $T_\mathrm{c}$ (Fig.~\ref{FWHM}). This confirms the collisionless nature of the BA mode at $k \sim k_\mathrm{F} / 2$ and identifies collisions with fermionic quasi-particles as the primary damping mechanism. Above $T_\mathrm{c}$ the width cannot be described by the collisionless model, but nor is it deeply hydrodynamic. To see this, we can compare our data to hydrodynamic theory based on viscous damping. Neglecting the contribution from thermal conductivity (discussed below), the damping rate is given by \cite{landau2013fluid}
\begin{align} 
    \hbar \Gamma / E_\mathrm{F} = \frac{8}{3} \left ( \frac{k}{k_\mathrm{F}} \right )^2 \frac{\eta}{n},
    \label{eta}
\end{align}
where $\eta$ is the shear viscosity. The blue dash-dotted curve in Fig.~\ref{FWHM}(b) is found using Eq.~\eqref{eta} with $\eta = 2.77 \hbar n (T/T_\mathrm{F})^{3/2}$, valid in the classical limit $T\gg T_\mathrm{F}$~\cite{Massignan2005,Bruun2005}. This is clearly well below the experimental data, which is not surprising given that $T < T_\mathrm{F}$. On the other hand, using the shear viscosity calculated from a self-consistent ${\mathcal T}$-matrix approach~\cite{Enss_AP2011} (red diamonds) yields a damping rate that appears consistent with the experimental data. However, this calculation ignores the damping contribution from thermal conductivity, which should be significant in the hydrodynamic regime~\cite{Braby2010, Patel_arXiv2019, Baird_PRL2019}. This further confirms that our experiments lie in the crossover between hydrodynamic and collisionless dynamics. For $\omega_0 \tau \sim 1$, there will be too few collisions to enable complete rethermalisation or momentum relaxation within one oscillation period of the sound wave. We thus expect both the viscous and thermal contributions to the damping to be lower than in the hydrodynamic limit as the efficiency of both mechanisms will be reduced.

A recent study of sound propagation in the unitary Fermi gas at long wavelengths, $k \lesssim k_\mathrm{F}/10$, measured the temperature dependence of sound diffusivity $D$ \cite{Patel_arXiv2019}. In the hydrodynamic limit, both the sound frequency $\omega_0$ and damping $\Gamma$ were found to evolve smoothly with temperature, including across $T_\mathrm{c}$. Below $T_\mathrm{c}$, a transition from hydrodynamic damping ($\Gamma \propto k^2$) to collisionless damping ($\Gamma \propto k$) was seen with increasing $k$ \cite{Patel_arXiv2019}, consistent with a model of sound attenuation in liquid helium \cite{Pethick_Physica1966}. Our data at larger $k$ lie outside the range of hydrodynamic $(\Gamma \propto k^2$) damping for all reported temperatures. Using the model of \cite{Pethick_Physica1966}, our damping rates below $T_c$ connect smoothly to those presented in \cite{Patel_arXiv2019} within experimental uncertainties. The jump in $\Gamma$ we observe at $T_\mathrm{c}$ highlights that the BA mode at $k \sim k_\mathrm{F}/2$ is a collisionless excitation driven by phase gradients and that the transition from hydrodynamic to collisionless sound takes place at different $k$ in the superfluid and normal phases. 






Finally, we compare the excitations in the unitary Fermi gas to those in liquid helium, which has been studied over several decades \cite{Cowley_CJP1971,Winterling_PRL1973,Tarvin_PRB1977,Maza_JPC1988,Stirling_PRB1990}. The brown stars in Fig. 3 show the scaled width (b), and quality factor $\omega_0/\Gamma$ (c) of the phonon mode in liquid helium~\cite{Cowley_CJP1971}, as a function of $T/T_\mathrm{c}$. Helium data were obtained via neutron scattering at a wavevector of $k \approx 0.4 \, \mathring{\rm{A}}^{-1}$, which lies approximately half way along the linear (phononic) branch of the dispersion curve, where the excitation energy is approximately half of the roton-gap energy. This is comparable to the phonon frequency in our measurements being roughly half of the pair-breaking threshold $2\Delta$. The qualitative similarity between these two systems as a function of the relative temperature is striking, despite the fact that one system is bosonic and the other fermionic. This intriguing similarity has attracted recent interest both theoretically \cite{Castin_PRL2017} and experimentally \cite{Patel_arXiv2019}.

\textit{Conclusion} We have studied sound propagation in a unitary Fermi gas as a function of temperature in the upper part of the phononic branch. The dominant excitation shows a strong temperature dependence and evolves from a collisionless BA mode driven by superfluid phase gradients below $T_\mathrm{c}$, to a strongly-damped mode, in the crossover between hydrodyamic and collisionless regimes above $T_\mathrm{c}$. While these different regimes are not distinguished by their frequency, the damping rates reveal the differences. Below $T_\mathrm{c}$, the damping is dominated by collisions with thermally excited quasiparticles and is well described by a QRPA theory, whereas above $T_\mathrm{c}$ the strong damping indicates that the mode lies in the crossover between the collisionless and hydrodynamic regimes. At even higher temperatures and excitation spectra approach the single-particle limit. Finally, we identify strong similarities in the temperature dependence of sound in the unitary Fermi gas and liquid helium at comparable momenta. 

We thank M.~Zwierlein, Y.~Castin and H.~Hu for stimulating discussions and J.~Denier for assistance on initial experiments. We acknowledge financial support from the Australian Research Council programs CE170100039, and from the Independent Research Fund Denmark - Natural Sciences via Grant No. DFF - 8021-00233B.

----------

\bibliographystyle{apsrev4-1}
\bibliography{ElementarExcitations}

\end{document}